\documentclass[conference]{IEEEtran}
\IEEEoverridecommandlockouts
\usepackage{cite}
\usepackage{amsmath,amssymb,amsfonts}
\usepackage{algorithmic}
\usepackage{graphicx}
\usepackage{hyperref}
\usepackage{pgfplots}
\pgfplotsset{compat=1.18}
\usepackage{textcomp}
\usepackage{xcolor}
\def\BibTeX{{\rm B\kern-.05em{\sc i\kern-.025em b}\kern-.08em
    T\kern-.1667em\lower.7ex\hbox{E}\kern-.125emX}}
\begin{document}

\title{Deep Contrastive Patch-Based Subspace Learning for Camera Image Signal Processing
}

\author{\IEEEauthorblockN{Yunhao Yang}
\IEEEauthorblockA{
\textit{University of Texas at Austin}\\
Austin, USA \\
yunhaoyang234@gmail.com}
\and
\IEEEauthorblockN{Yi Wang}
\IEEEauthorblockA{
\textit{University of Texas at Austin}\\
Austin, USA \\
panzer.wy@utexas.edu}
\and
\IEEEauthorblockN{Chandrajit Bajaj}
\IEEEauthorblockA{
\textit{University of Texas at Austin}\\
Austin, USA \\
bajaj@cs.utexas.edu}
}

\maketitle

\begin{abstract}
Camera Image Signal Processing(ISP) pipelines can get appealing results in different image signal processing tasks. Nonetheless, the majority of these methods, including those employing an encoder-decoder deep architecture for the task, typically utilize a uniform filter applied consistently across the entire image.
However, it is natural to view a camera image as  heterogeneous, as the color intensity and the artificial noise are distributed vastly differently, even across the two-dimensional domain of a single image. 
Varied Moire ringing, motion blur, color-bleaching, or lens-based projection distortions can all potentially lead to a heterogeneous image artifact filtering problem. 
In this paper, we present a specific patch-based, local subspace deep neural network that improves Camera ISP to be robust to heterogeneous artifacts (especially image denoising). We call our three-fold deep-trained model the Patch Subspace Learning Autoencoder (PSL-AE)
\footnote{Our source code is publicly available at \href{https://github.com/CVC-Lab/Patch-Subspace-Learning-Autoencoder}{https://github.com/CVC-Lab/Patch-Subspace-Learning-Autoencoder}}. 
The PSL-AE model does not make assumptions regarding uniform levels of image distortion.
Instead, it first encodes patches extracted from noisy and clean image pairs, with different artifact types or distortion levels, by contrastive learning. 
Then, the patches of each image are encoded into corresponding soft clusters within their suitable latent sub-space, utilizing a prior mixture model. Furthermore, the decoders undergo training in an unsupervised manner, specifically trained for the image patches present in each cluster. The experiments highlight the adaptability and efficacy through enhanced heterogeneous filtering, both from synthesized artifacts but also realistic SIDD image pairs.
\end{abstract}


\section{Introduction}
In a Camera's Image and Signal Processing (ISP) pipeline, specialized digital signal processors are important in converting RAW images from the camera's digital sensors into conventional RGB or JPEG images. To address various image artifacts (distortions) during this conversion process, camera manufacturers consistently seek and necessitate the development of advanced filters as integral components of their camera ISP systems.
Such filters includes demosaicing \cite{li2008image}, deblurring \cite{baek2008noise}, color correction \cite{kwok2013simultaneous}, etc. 
However, with the evolution of camera hardware and the increment in image resolution, traditional filters have limited capabilities in resolving artifacts or distortions.

Algorithms and models with deep learning techniques are increasingly supplanting image and signal processing techniques on traditional computational photography tasks. For instance, neural networks with hierarchical structures that encode low-level details display significant performance on image deblurring (e.g., \cite{abdelhamed2019ntire,valsesia2020deep})  and deblurring (e.g., \cite{kupyn2019deblurgan,suin2020spatially}) tasks. Nevertheless, most of these algorithms or models assume there is a universal artifact. They have limited capabilities to capture heterogeneous artifacts caused by complicated environmental settings in the real world. Due to sensor limitations, environment changes, or post-processing like image compression, images may contain non-universal, heterogeneous artifacts across different locations. Current models that assume universal artifacts are not suitable for removing such heterogeneous artifacts.

In this paper, our goal is to produce a high-quality deblurred RGB image from noisy RGB images, where we do not rely on uniform artifacts assumption. Moreover, we do not assume spatial information; rather, we treat images as unions of patches. There are several works that focus on learning latent representation in pixel or patch level \cite{pix2pix2017,gupta2020patchvae}. Our method is different from these previous works. We encode each patch (including noisy-clean contrastive pairs) into the target underlying latent space and learn the most appropriate subspace that distinguishes noisy parts with different levels and types. When latent space gets projected back to the image space, the network can generate heterogeneous patch results by passing them into different decoders. We assemble these patches and obtain the deblurred result under a wider range of artifacts or corruption assumptions. Unlike previous work, our network is relatively simple yet powerful under multiple source noise denoising tasks (c.f. synthesized CelebA dataset). Moreover, to show the potential of our approach, we not only test on synthetic datasets but also conduct experiments over a realistic, challenging dataset (i.e., SIDD dataset).

\section{Related Works}

\paragraph{Image Denoising}
Noise can be introduced to images during the image acquisition process, transmission, or storage. Numerous approaches are proposed to remove different types of noise within in image domain and seldom focus on non-local denoising. For instance, BM3D \cite{dabov2007image} applies the idea of group filtering via transforming stacking of image patches in the 3D domain with sparse representation. Another work \cite{ghimpecteanu2016local} focuses on non-local smoothing under lower complexity compared with BM3D. 
After the deep learning approach is explored, deep neural networks, such as convolutional neural networks and generative adversarial networks, significantly improve the denoising performance compared to the classical methods \cite{guo2018convolutional,xu2020noisy, moran2020noisier2noise}. Nonetheless, owing to the costly nature of the capture process and the inherent diversity of noise in large-scale, realistic photographs, the majority of previous research has been confined to investigations within the synthesized domain. For example, as demonstrated by earlier benchmarks, the widely utilized Additive White Gaussian Noise (AWGN) model is inadequate for effectively eliminating noise from actual images \cite{plotz2017benchmarking}. 

To augment the missing pair of a realistic noisy, and clean image, self-supervision brings more attention to the field. The idea of self-supervision allows the network to learn from images even without specifying clean/noisy pairing. Earlier works N2V \cite{krull2019noise2void}, N2S \cite{batson2019noise2self}, and N2N \cite{moran2020noisier2noise} all follow the spirit of learning noise patterns wildly without specifying noisy-clean pairs when performing training. Recent works are learning under more versatile conditions, namely more specific self-supervised task design \cite{Liu2021RelightingII,Khademi2021SelfSupervisedPD}. Although our idea introduces matched noisy/clean image pairs as a prior knowledge of class labeling, we show that, with a contrastive learning pipeline, our encoder achieves better performance compared with previous self-supervised neural networks if we no longer assume a single noise additive from the clean image.

\paragraph{Deep Learning Within Camera ISP Scope}
Many works explore applying deep learning in various tasks related to Camera ISP, such as color demosaicing (e.g., \cite{khashabi2014joint}) and image denoising (e.g., \cite{tian2020deep}).
CycleISP \cite{zamir2020cycleisp} develops a generative model to generate synthesized realistic images for image denoising. Moreover, another work \cite{schwartz2018deepisp} can perform image denoising by employing deep neural networks derived from the original sampling of sensors. 
A more recent work, PyNET \cite{ignatov2020replacing}, introduces a neural network containing give parallel learning levels to supplant the entirety of the Camera ISP pipeline.

\paragraph{Latent Subspace Learning}
Our primary contribution involves the development of a method for learning a latent encoding that optimizes the utilization of self-similarity present in natural images by probing subspaces. This research encompasses at least two distinct workflow branches. The first branch emphasizes the direct disentanglement of latent space clustering through variational inference. Existing works 
\cite{bouchacourt2018multi,ding2019clustering, yang2022training} develop methods to divide the latent space into subspaces or hierarchical spaces to learn better latent-space projections.
Another branch is the integration of deep learning models with classical machine learning algorithms. 
These works apply clustering algorithms (e.g., Soft K-means Clustering \cite{jabi2019deep}), mixture models (e.g., Gaussian Mixture Model VAE \cite{dilokthanakul2016deep}), or direct subspace clustering VAE \cite{klys2018learning} within the latent space.

Each of these distinguished models successfully acquires knowledge from visual recognition and classification tasks through the employment of a generative model. In contrast, we do not introduce additional noise in the latent space, but we keep the idea of latent space ``pseudo-labeling" to feed image patches into different filters. 
\section{Patch-Based Subspace Learning Autoencoder}

In this section, we describe our Patch Subspace Learning Autoencoder(PSL-AE) to encode image patches and classify them into unsupervised latent subspaces. 
Our objective is resolving image artifacts, specifically image denoising. 
One key underlying assumption in our algorithm is that the additive noise model is a mixture model containing different types of noise. Namely, 
\begin{equation}
    f(I_{obs}) = I_{gt} + \sum_{s=1}^S \Sigma_s \odot M_s.
    \label{eq:model:mixture:noise}
\end{equation}

In \eqref{eq:model:mixture:noise}, $f$ is the target mapping function we shall learn to recover ground truth $I_{gt}\in \mathbb{H\times W\times C}$, i.e., clean images, from the noisy observation $I_{obs}\in \mathbb{H\times W\times C}$. It shall close or be identical to the identity function when we are performing sRGB to sRGB image denoising. The noise $\Sigma_s\in \mathbb{H\times W\times C}$ and the mask $M_s\in \mathbb{H\times W\times C}$ are independent and unknown to the model, with $\odot$ refers to the element-wise product. $S$ denotes the number of noise types, which is a hyper-parameter under most scenarios, and our model does not rely on the knowledge of $S$. Moreover, we do not assume the underlying distribution $S$ types of noise.

\begin{figure}[t]
\begin{center}
    \includegraphics[width=\linewidth]{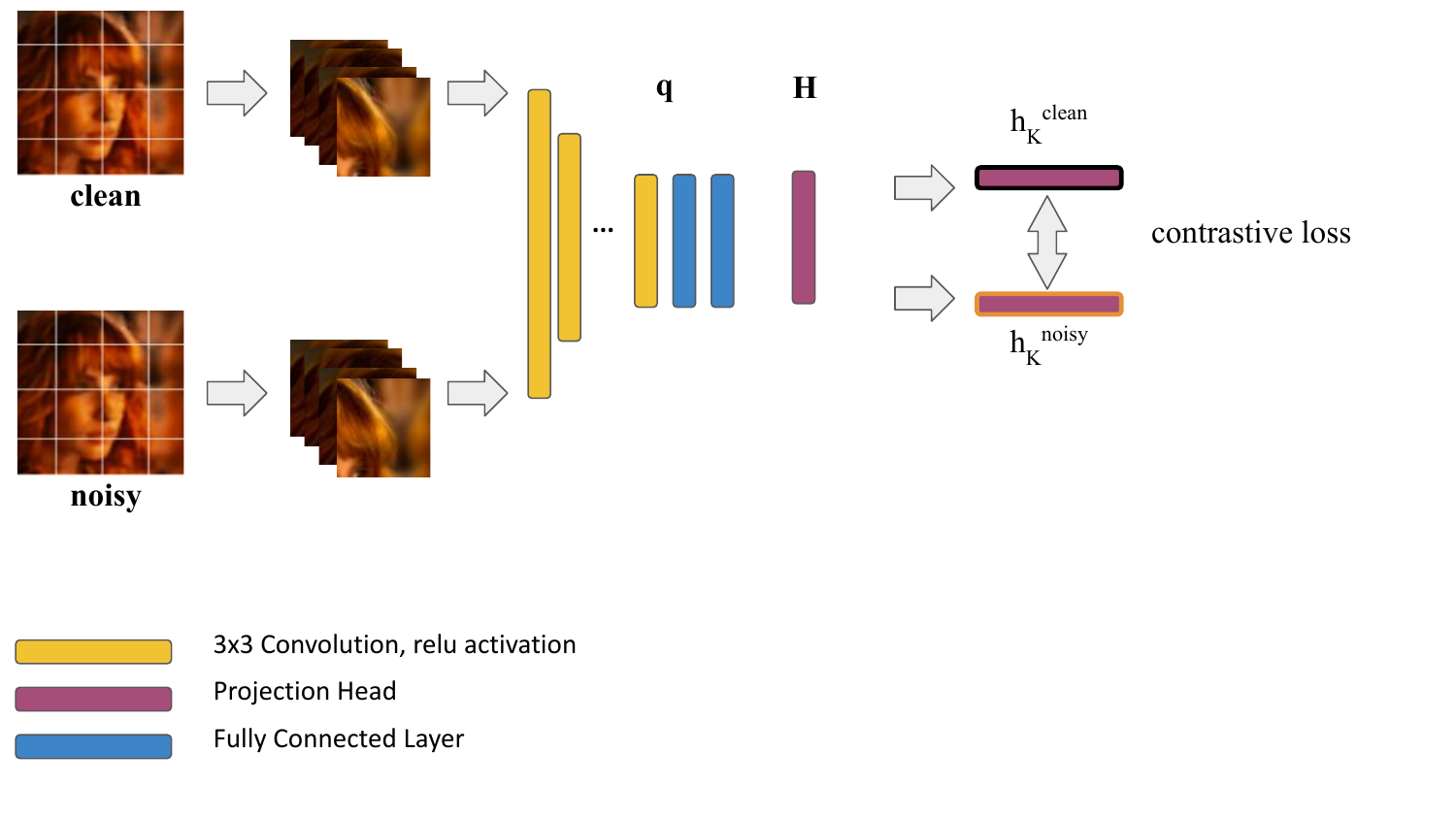}
\end{center}
\caption{In the pretraining stage, we add a projection head $H$ to the encoder $q$ to adapt pseudo-labels. We pair the clean patch and its corresponding noisy patch to compute the contrastive loss, which means each clean-noisy patch pair consists of a unique pseudo-label. Then, we train the encoder to minimize the contrastive loss. In the training and validation stage, we remove the projection head of the pretrained encoder and use it as the encoder of our PSL-AE.}
\label{fig:struct:contrastive}
\end{figure}

\textit{Overall Pipeline of PSL-AE}: 
Our proposed PSL-AE is an autoencoder with a single encoder and multiple decoders. For every input image $I_{obs}$, we first split the input image into $D$ by $D$ patches $P_n\in\mathbb{R}^{D\times D \times C}$ with overlap allowed. For every patch $P_n$ we assign a location mask matrix $H_n$ and denote $P_n\overset{def}{=} I_{obs}\odot H_n$. Even though we cannot probe the magnitude of the mixed artifacts, we assume the artifacts within the patch can be approximated with a single dominant denoise decoder:
\vspace{-1.0em}
\begin{align}
    f(P_n) = f(I_{obs} \odot H_n) &=  I_{gt}\odot H_n + \sum_{s=1}^S  \Sigma_{s} \odot M_{s} \odot H_n ,\nonumber \\
    & \approx (I_{gt} + \Sigma_i \odot M_i) \odot H_n.
\label{eq:model:mixture:patch}
\end{align}

Our PSL-AE network learns to map from image patches $P_n$ to $I_{gt}\odot H_n$. Our encoder $q(P_n|\vec{\theta})$, with network parameter $\vec{\theta}$, takes a batch of image patches as the input. 
It will generate a latent vector $\vec{z}_n$ and a soft-labeling vector $\vec{y}_n$ as the latent subspace indicator of each patch. Then, based on the latent space clustering result, we set up $k$ decoders $r_i(\vec{z}_n|\vec{\psi}_i), i=1,2\ldots,k,$ with parameter $\vec{\psi}_i$. Each patch's latent encoding $\vec{z}_n$  is fed into $y_n$-th decoder based on soft-labeling vector $\vec{y}_n$, i.e. $y_n\overset{def}{=}\arg\max_i~(\vec{y}_n)_i$. In our settings, all $k$ decoders have the same architecture, but their parameters are independent. These decoders receive different numbers of patches within one batch and are trained simultaneously. The output $r_{y_n}(q(P_n|\vec{\theta})|\vec{\psi})$ is the reconstructed block that approximates $I_{gt}\odot H_n$. Although we have imbalanced training data in each batch sent to different decoders, and there will be increasing parameters as $k$ goes up, our experiment results demonstrate that the overall scheme is efficient and does not deteriorate the network performance.

\subsection{Patch-Based Self-Supervised Pre-training}
Our patch-based encoder plays an important role in projecting noisy and clean image patterns into separable subspace feature vectors. Henceforth, we are in particular interest in learning from noisy to clean patch-wise. It has been proved that imposing  self-supervision \cite{xu2020noisy} tasks can improve the generalizability and performance of a neural network. We adopt the same methodology presented in  SimCLR \cite{chen2020simple} and propose to learn noisy-clean contrastive patches from source data, which served as a preprocessing step. We first collect a set of images and divide them into patches, denoting each patch $P_n$. Then, we artificially generate noisy patches from the clean patches; we denote $f(P_n)$ as the noisy patch generated from $P_n$. Note that $f$ is defined by us, which means the noise type and hyper-parameters are known.

The base encoder $q$ takes the patches as input and maps them on the latent space, then a projection head $H$ is connected to the base encoder. We use the patch location as the pseudo-label. Hence a clean patch $P_n$ and its corresponding noisy patch $f(P_n)$ will be assigned the same pseudo-label $n$. The pseudo label refers to a location embedding token that serves as the self-supervised class labeling. The contrastive loss is maximizing the agreement between the clean patches and their corresponding noisy patches. Each projection $h_k^{clean} = H(q(P_k))$ for the clean patch and projection $h_k^{noisy} = H(q(f(P_k)))$ for the noisy patch are used as a pair to compute the contrastive loss.

The similarity metric we use is:
\begin{equation}
    sim(u, v) := \frac{u^T v}{||u|| \cdot ||v||}.
\end{equation}

We apply the NT-Xent \cite{chen2020simple} (normalized
temperature-scaled cross-entropy loss) as our contrastive loss function:
\begin{equation}
    \mathcal{CL}_{i, j} = - \log \frac{\exp( sim(\vec{h}_i, \vec{h}_j) /\tau )}{\sum_{k=1}^{2N}  \mathbf{1}_{k \ne i} \exp (sim(\vec{h}_i, \vec{h}_k) /\tau) }.
\end{equation}
where $N$ is the batch size, $\mathbf{1}$ is the indicator, $h$ is the output of the projection head, and $\tau$ is the temperature variable that defined by us. The NT-Xent incorporates a temperature scaling parameter to adjust the probability distribution produced by the model, ultimately enhancing calibration and generalization performance. This self-supervised pre-training procedure is demonstrated in Figure \ref{fig:struct:contrastive}.

\begin{figure}[t]
\begin{center}
    \includegraphics[width=\linewidth]{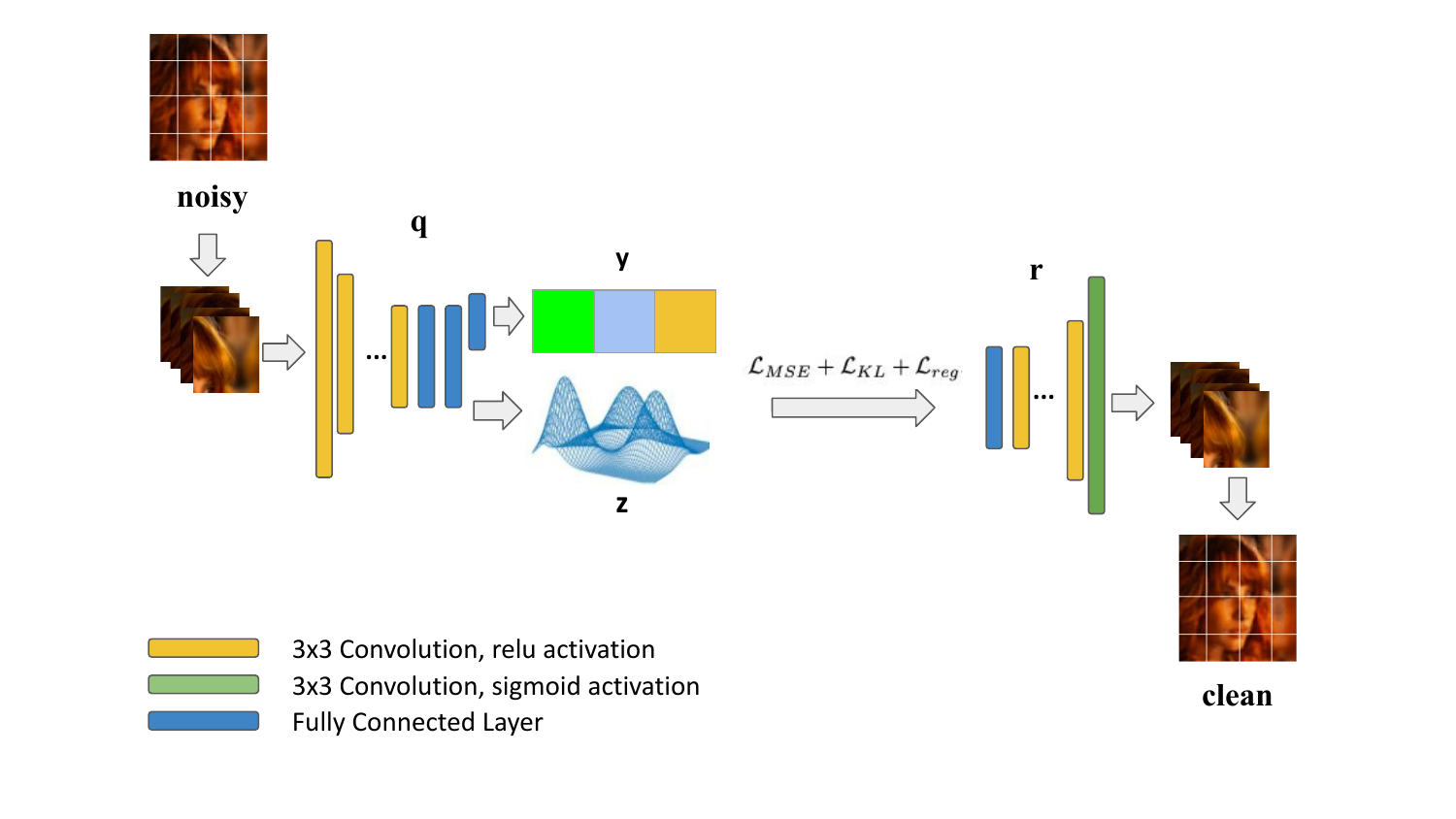}
\end{center}
\caption{In the training stage, we divided the given image into patches and took each of them as the input of the encoder. In latent space, we generate the codex $\vec{z}$ to preserve features for reconstruction and soft-categorical vector $\vec{y}$ to identify the subspace that each patch belongs to. We build and train the encoder to achieve the best latent encoding with a latent vector maximally separating patches into subspaces and a dummy decoder.}
\label{fig:struct:enc}
\end{figure}

\subsection{Multi-filter Denoising With Multi-Decoder}

\begin{figure*}[t]
\begin{center}
    \includegraphics[width=0.75\linewidth]{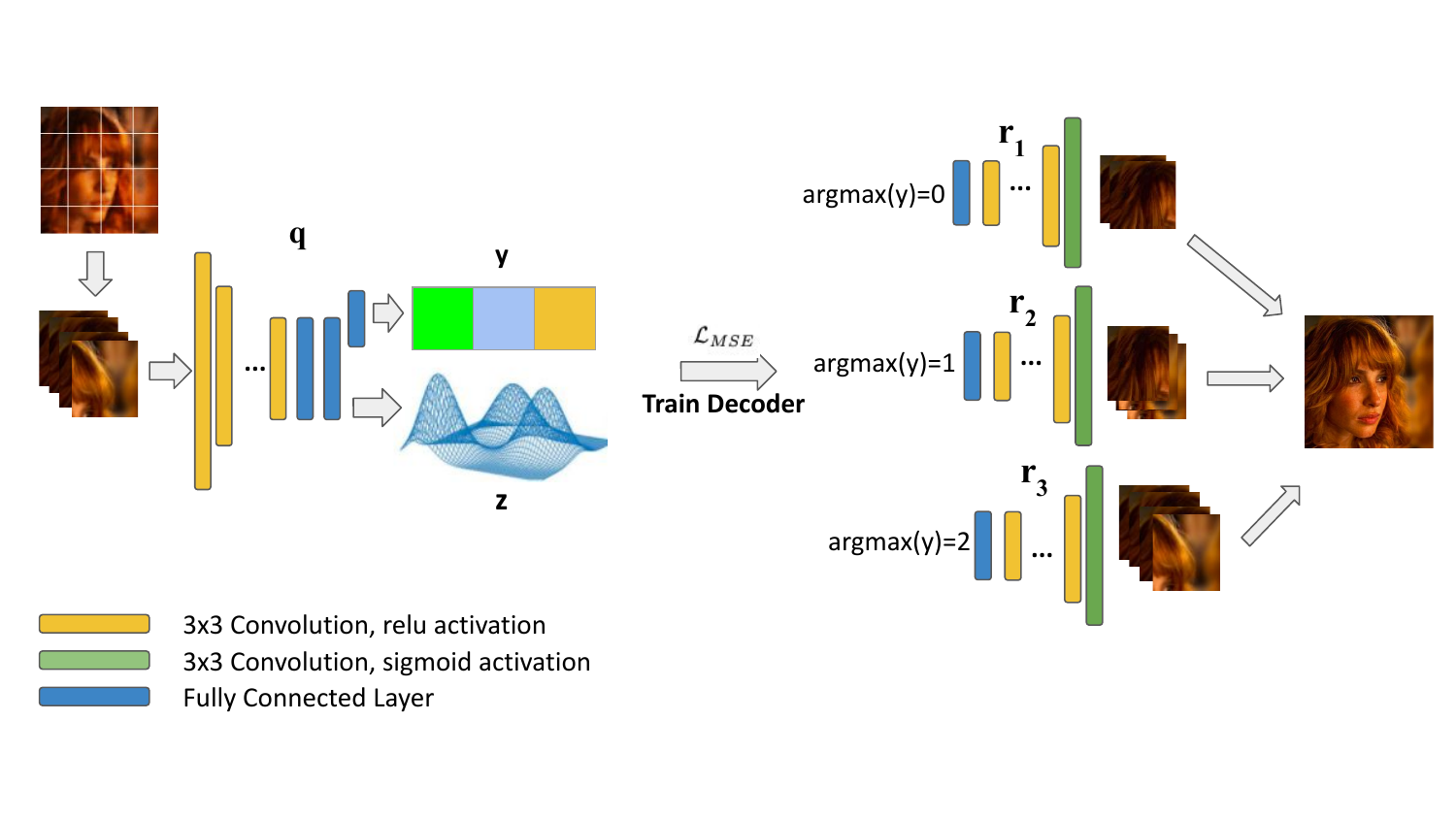}
\end{center}
\caption{In latent space, the encoder generates the codex $\vec{z}$ and soft-categorical vector $\vec{y}$ to identify the subspace that each patch belongs to, based on its noise level and similarity with other patches. We take $\vec{z}$ and send it to different decoders based on labeling $\vec{y}$. The final result of a reconstructed image is obtained via assembling all patches of the image.}
\label{fig:struct:dec}
\end{figure*}

Our PSL-AE consists of an encoder- initialized from the contrastive learning encoder, a dummy decoder to help reconstruct the patches and computing losses, a set of decoders to reconstruct vectors from different subspaces. The training procedure can be divided into two steps: we first train the encoder and dummy decoder to formalize latent subspaces; second, we fix the encoder and use the data from different subspaces to train decoders, where each subspace corresponds to a unique decoder.

\paragraph{Training the encoder and dummy decoder}
Our two-stage training optimizes different loss functions. The weights in the encoder are initialized from the contrastive pretrained encoder. For encoder training, our encoder training loss $\mathcal{L}_{enc}$ consists of an evidence lower bound (ELBO) loss, including the reconstruction loss and the KL loss, and an L2 regularization loss.
\begin{equation}
        \mathcal{L}_{enc}= \mathcal{L}_{MSE}+ \mathcal{L}_{KL}+ \mathcal{L}_{reg},
    \label{eq:loss:enc}
\end{equation}
where
{\small
\begin{align}
    \mathcal{L}_{MSE} =& \sum_{n=1}^N \|r(q(P_n|\vec{\theta})|\vec{\psi})- P_n \|_F^2, \\
      \mathcal{L}_{KL} = &\quad - \lambda_{y}\sum_{n=1}^N \sum_{s=1}^S (\hat{\vec{y}}_n)_s\log \frac{(\hat{\vec{y}}_n)_s}{\vec{y}_s},  \label{eq:loss:enc:KL} \\
    \mathcal{L}_{reg} =& \lambda_{reg}(\|\vec{\theta}\|_2^2 + \|\vec{\psi}\|_2^2). \label{eq:loss:enc:reg}
\end{align}
}

The decoder $r_0$ is a "dummy" decoder with parameter $\vec{\psi}$, where we rely on its reconstruction result in the original patch space. The prior distribution $y$ is a categorical distribution while the latent estimation is the soft-labeling vector $\vec{y}_n$, thereby forming the second half of (\ref{eq:loss:enc:KL}). $\lambda_y$ is the weight parameter and $\lambda_{reg}$ controls the additional regularization. Note that since the artifact model is blind to us, we cannot get the true prior $p(y)$. Instead, we predefine the vector $\vec{y}$ as the fixed discrete categorical distribution. Without further specification, the prior is assumed to be uniform:  $\vec{y}=(\frac{1}{S},\frac{1}{S},\cdots,\frac{1}{S})^T$.
We present this procedure in Figure \ref{fig:struct:enc} and address the choice of other hyper-parameters for different datasets respectively in Section \ref{sec:experiment}.

\paragraph{Training multiple decoders}
For decoder training, the loss function $\mathcal{L}_{dec}$ no longer minimizes the KL divergence as the latent encoding is fixed, so only the distribution of patches is optimized:
\begin{equation}
        \mathcal{L}_{dec}= \mathcal{L}_{MSE}+ \mathcal{L}_{reg},
    \label{eq:loss:dec}
\end{equation}
where
\vspace{-0.5em}
{\small
\begin{align}
    \mathcal{L}_{MSE} =& \sum_{n=1}^N \|r_{\color{blue}y_n}(q(P_n|\vec{\theta})|\vec{\psi}_{\color{blue}y_n})- P_n \|_F^2, \\
    \mathcal{L}_{reg} =& \lambda_{reg} \sum_{s=1}^S \|\vec{\psi}_s\|_2^2. \label{eq:loss:dec:reg}
\end{align}
}%
We fix the encoder $q$ during the entire training process, thereby fixing the labeling $y_n$ for every input patch $P_n$. The separation of training decoders from the encoder allows different decoders to learn to distinguish features within grouped patches, including similar patterns. The regularization weight $\lambda_{reg}$ in (\ref{eq:loss:dec:reg}) is the same as the one in (\ref{eq:loss:enc:reg}), and it is further specified in our experimental settings. The multi-decoder training is demonstrated in Figure \ref{fig:struct:dec}.

\paragraph{Patch Assembly}
We merge the reconstructed blocks outputted by the decoder into the original-size image. However, we can observe a clear boundary between the neighbor blocks when we merge them together, and such boundaries lead to new artifacts. To avoid causing new artifacts during the merging process, we concatenate the blocks such that each two neighbor blocks share a small overlapping region. 
The output sRGB values in overlapping regions are filtered and smoothed out via linear interpolation between two overlapped patches.
\section{Experiments}
\label{sec:experiment}

We justify the performance of our PSL-AE over two datasets: CelebFaces Attributes (CelebA) Dataset \cite{liu2015faceattributes} with synthesized noise and Smartphone Image Denoising Dataset (SIDD) \cite{abdelhamed2018high}. 
Among the two datasets, Synthesized Noisy CelebA has a mixture of artifacts that exactly matches our model in (\ref{eq:model:mixture:noise}).
The SIDD consists of images from smartphone cameras with realistic artifacts. We expect a performance enhancement of our multi-decoder architecture on resolving realistic artifacts, compared to a single decoder convolutional autoencoder, referred to as \textit{Single Decoder Conv}.
\begin{figure}[t]
\begin{center}
    \includegraphics[width=\linewidth]{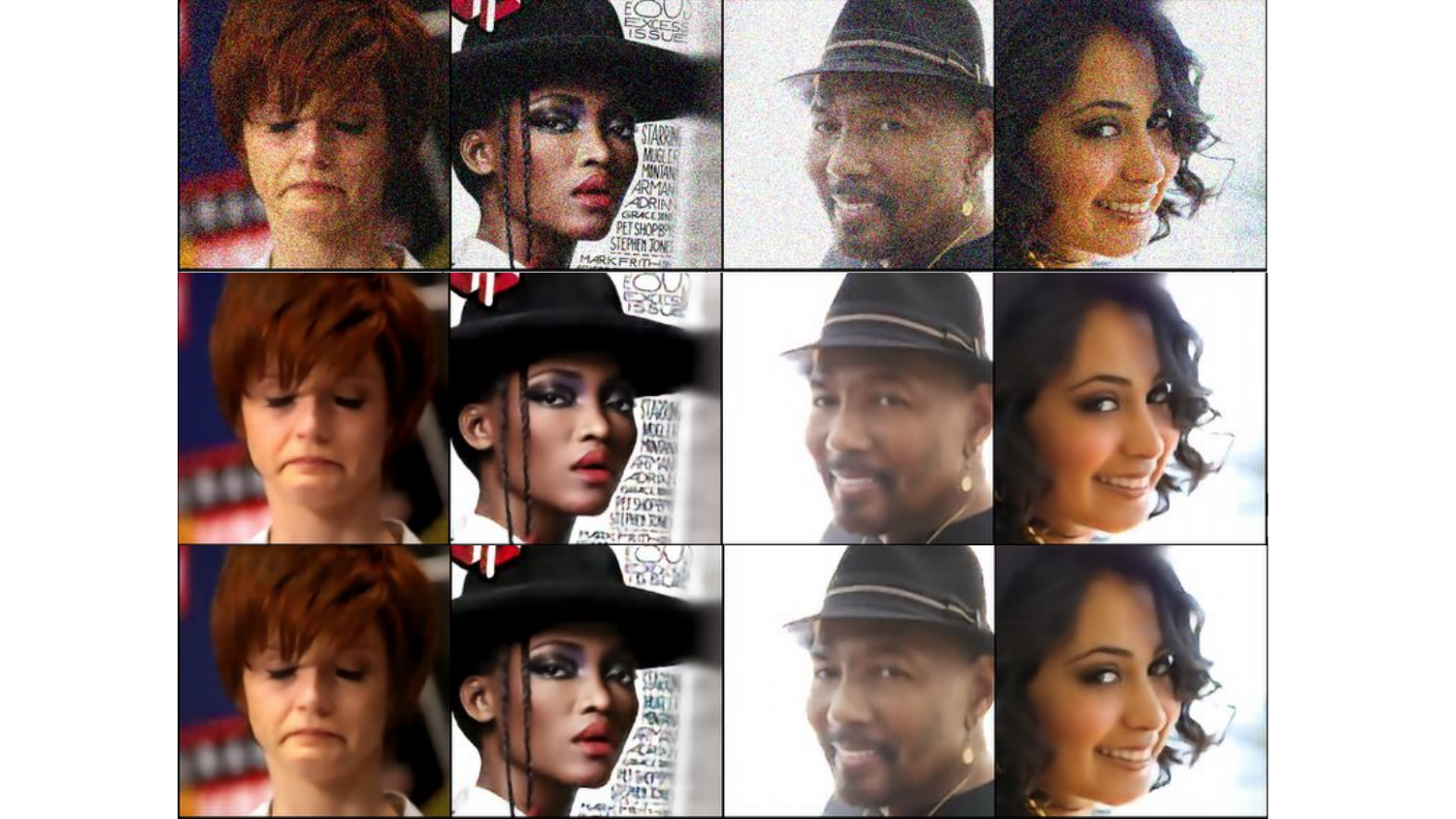}
\end{center}
\caption{CelebA Global Noise Denoising Results: images from the top to bottom rows are the noisy images with Gaussian noise $\sigma=0.2$, denoising results from a single-decoder convolutional autoencoder, denoising results from a 2-decoder PSL-AE with contrastive pre-training, respectively.
We are able to show that the denoising results from both the convolutional autoencoder and our PSL-AE are close to the ground truth, and our method performs insignificantly better than the convolutional autoencoder.}
\label{fig:celeba:results}
\end{figure}
\subsection{Our Networks with CelebA Patches}
\label{sec:celeba}

\paragraph{Dataset Construction}
We adopt a subset of the Large-scale CelebFaces Attributes (CelebA) Dataset, CelebA\_HQ/256 dataset, for our first set of empirical results.
CelebA\_HQ/256 consists of 2250 training images and 11250 validation images with size $256 \times 256$ pixels. 
We manually generate artifacts on the images to train and evaluate our network. Our PSL-AE is implemented using Keras\cite{chollet2015keras} and Tensorflow\cite{tensorflow2015-whitepaper}. The detailed implementation of our network is presented in the Supplementary.

\paragraph{Global Noise Generation}
We apply a global artifacts generator to the CelebA dataset to obtain a dataset with globally distributed noise, referred to as the synthesized global noise/artifact dataset. 
We generate Gaussian noise with $\sigma=0.2$ on these images using OpenCV\cite{opencv_library} library. The Gaussian noise is uniformly distributed in the images. The noisy images have an average PSNR of $17.78$.

\paragraph{Our denoising result on CelebA dataset with synthesized global artifacts}

\begin{table}[t]
\caption{CelebA Results Comparison on Synthesized Global Artifacts}
\begin{center}
 \begin{tabular}{||c c c c||} 
 \hline
  & PSNR & SSIM & UQI \\ [0.5ex] 
 \hline
 Image with Artifacts & 17.78 & 0.6038 & 0.7527 \\
 
 N2V\cite{krull2019noise2void}& 21.66 & 0.7242 & 0.9249 \\
 
 N2N\cite{moran2020noisier2noise}& 26.60 & / & / \\
 
 VAE\cite{jiwoong2017DVAE} & 26.81 & 0.7621 & 0.9604 \\
 
 RSE-RL\cite{ch2021reinforcement} & 29.03 & 0.8339 & 0.9731\\
 
 UNet\cite{ronneberger2015unet} & 28.74 & 0.8579 & 0.9731 \\
 
 Single Decoder Conv & 28.53 & 0.8427 & 0.9733 \\
 
 PSL-AE (2 decoders) & \textbf{29.59} & \textbf{0.8638} & \textbf{0.9741} \\ [1ex] 
 \hline
\end{tabular}
\end{center}
\label{tab:celeba1}
\end{table}

The evaluation metrics we use in this work are PSNR, SSIM\cite{wang2004image}, and UQI\cite{wang2002universal} scores. We use the convolutional autoencoder (Single Decoder Conv), variational autoencoder (VAE), Noise2Void (N2V), and the pre-trained Noise2Noise (N2N) model for Gaussian noise as the benchmarks. 
While deploying N2V, we divide 400 CelebA synthesized noisy images into $400 \times 128$ patches, where the size of each patch is $16\times 16\times 3$.
And we tested N2V on 1575 images with synthesized noise. 
In addition, we apply one of the state-of-the-art methods, UNet \cite{ronneberger2015unet}, on our synthesized dataset to compare with our results. We train the UNet under identical settings with our PSL-AE.

Table \ref{tab:celeba1} provides the baseline results and our denoising results. We divide each of the 2250 training images into 441 $16\times 16\times 3$ patches and feed them into the network. A visualization of the results is presented in Figure \ref{fig:celeba:results}, where we can see our PSL-AE is slightly better than the convolutional autoencoder.

\paragraph{Heterogeneous Noise Generation}
We apply a heterogeneous artifact generator to images from the CelebA dataset, referred to as the \textit{heterogeneous noisy CelebA dataset}. Three types of artifacts - Gaussian, Poisson, Salt and Pepper noise - are randomly generated on these images utilizing OpenCV\cite{opencv_library}. Each image contains three types of artifacts in different regions, while the regions can be overlapped with each other. The noisy images have an average PSNR of $19.35$. The noisy and ground truth images are divided into $16 \times 16$ pixels patches, with 4 pixels overlapping with the surrounding patches.

\paragraph{Our denoising result on CelebA dataset with synthesized heterogeneous artifacts}

\begin{table}[!htbp]
\caption{CelebA Results Comparison on Synthesized Heterogeneous Artifacts}
\begin{center}
 \begin{tabular}{||c c c c||} 
 \hline
  & PSNR & SSIM & UQI \\ [0.5ex] 
 \hline
 Image with Artifacts & 19.35 & 0.6582 & 0.7956 \\
 
 BM3D & 25.13 & 0.7639 & 0.9428 \\
 
 N2V\cite{krull2019noise2void}& 23.21 & / & / \\
 
 N2N\cite{moran2020noisier2noise}& 26.94 & / & / \\
 
 VAE\cite{jiwoong2017DVAE} & 27.37 & 0.7848 & 0.9611 \\
 
 UNet\cite{ronneberger2015unet} & 30.85 & 0.8759 & 0.9673 \\
 
 
 Single Decoder Conv & 31.02 & 0.8831 & 0.9769 \\
 
 PSL-AE (4 decoders) & \textbf{31.74} & \textbf{0.8996} & \textbf{0.9802} \\ [1ex]
 \hline
\end{tabular}
\end{center}
\label{tab:celeba2}
\end{table}
Table \ref{tab:celeba2} provides the results on the synthesized heterogeneous noise dataset. The training and testing procedures and the baseline models are identical to the setups on the synthesized global noise dataset.
The results indicate that PSL-AE significantly enhanced the denoising outcomes, according to all three metrics. 
\begin{figure}[t]
\begin{center}
    \includegraphics[width=\linewidth]{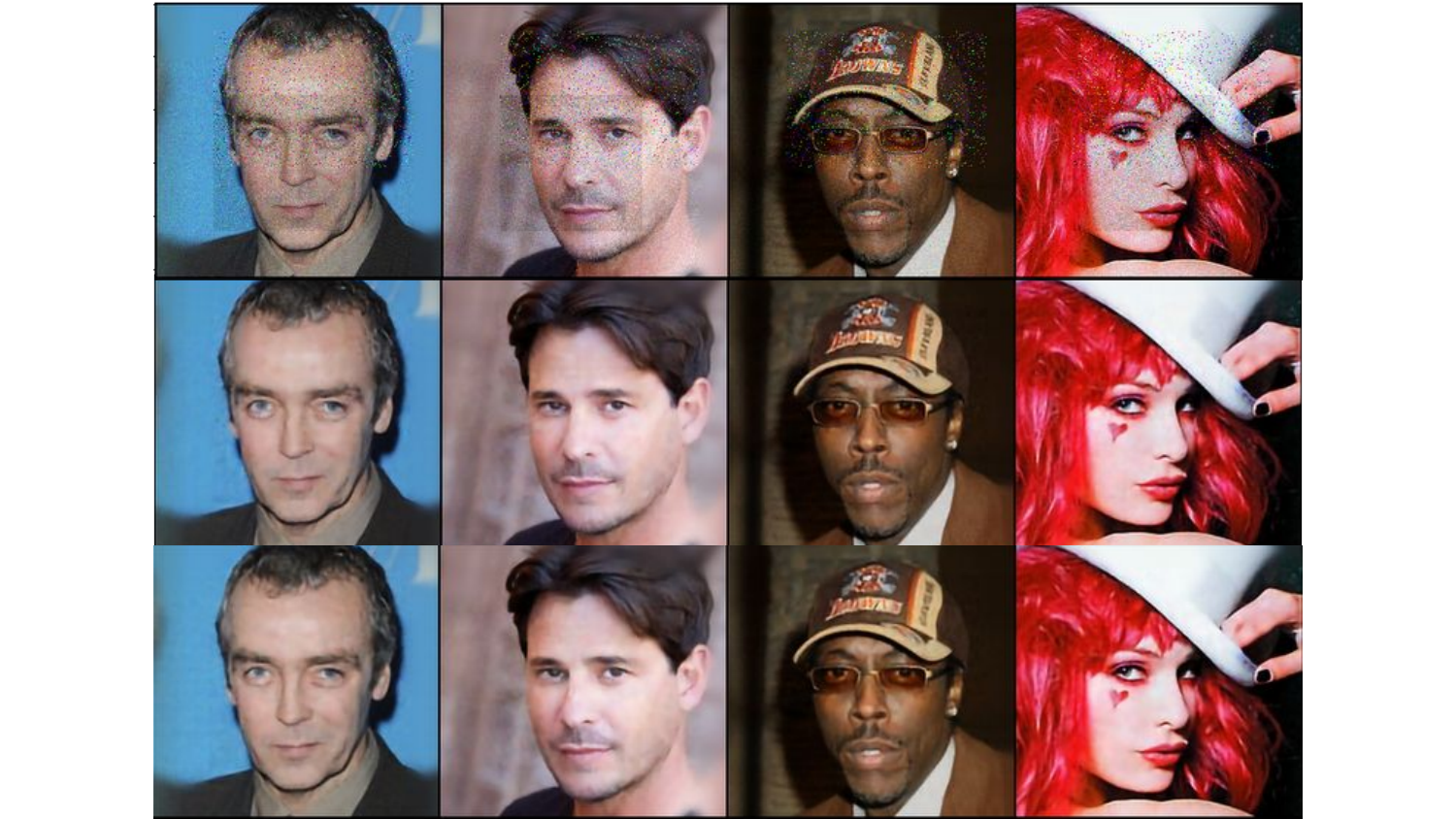}
\end{center}
\caption{CelebA Noise Denoising Results on Heterogeneous Artifacts. The first row shows the noisy images, the second row shows the denoising results from a single-decoder convolutional autoencoder, and the third row shows the results of our four-decoder PSL-AE. We can observe the enhancement of our PSL-AE compared to the convolutional autoencoder.}
\label{fig:celeba:results3}
\end{figure}

\begin{figure}[t]
\begin{center}
    \includegraphics[width=0.8\linewidth]{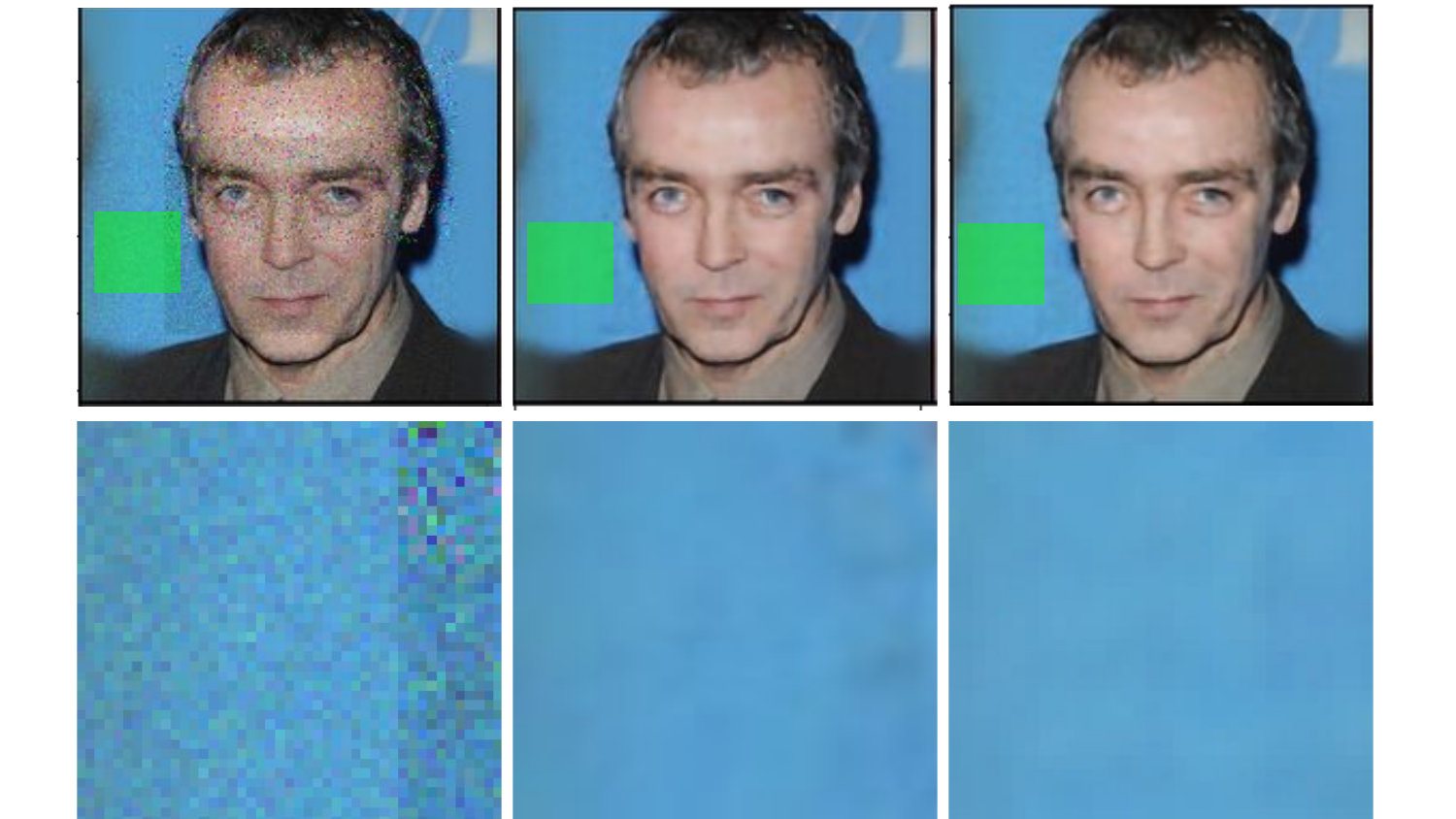}
\end{center}
\caption{Zoomed Denoising Results on Heterogeneous Artifacts: the left column is the noisy image, the middle column is the denoising result by the single decoder convolutional autoencoder, and the right column is the denoising result by our four-decoder PSL-AE. We zoom in on the green area to observe more details. The denoising result from the convolutional autoencoder still contains artifacts near the boundary of two different noise regions. By contrast, our PSL-AE utilizes multiple decoders to focus on different types of noises and thus resolve the artifacts on the boundary.}
\label{fig:celeba:results4}
\end{figure}

We conduct a set of experiments to explore the relationship between the denoising quality and the number of decoders in our PSL-AE. Experimental settings are described in Section \ref{sec:ablation}. We present our result in Figure \ref{fig:celeba_num_filter}. Among all the results, 4-decoder PSL-AE achieves the best performance. The performance matches our settings of three different types of artifacts plus one filter that pass the clean patches. However, there is no significant improvement in terms of the quality metrics when the number of decoders is greater than 4. The patches are clustered based on their projections in the latent space. It implies that patches assigned to the same cluster are likely to contain similar features. Each decoder is trained to reconstruct data with similar features. Conceptually, more decoders are likely to result in better performance due to a more detailed feature separation, but more decoders increase the possibility of assigning patches to the wrong cluster. A visualization of our reconstructed image results can be found in Figure \ref{fig:celeba:results3}. Figure \ref{fig:celeba:results4} presents a better visualization of why our multi-decoder PSL-AE outperforms the single-decoder autoencoder.

\subsubsection{Ablation Study}
\label{sec:ablation}
To demonstrate the effectiveness of our current setups, we conduct a set of ablation studies on the synthesized CelebA dataset and compare the results.

First, we explore how the number of decoders of our PSL-AE affects the denoising quality. We perform two sets of experiments to observe the models' outcomes on different numbers of decoders:
First, we explore how the number of decoders of our PSL-AE affects the denoising quality. We perform two sets of experiments to observe the models' outcomes on different numbers of decoders:
\begin{itemize}
    \item In the first set of experiments, we simply train and evaluate the models with the number of decoders from 1 to 8. Note that the structures of decoders are identical, hence the network complexity (evaluated by flops) is increasing while the number of decoders increases.
    \item In the second set of experiments, we manually control the network complexity by modifying the convolutional layers in the decoder. We adjust the kernel size of every convolutional layer and monitor the number of flops of the network. Suppose the kernel size of single decoder PSL-AE is $K$, then we can set the kernel size of $n$-decoder PSL-AE to $\frac{2K}{n+1}$ to ensure the complexities are on the same scale.
\end{itemize}
The results are presented in Figure \ref{fig:celeba_num_filter}.

\begin{figure}[!ht]
    \centering
\begin{tikzpicture}

\definecolor{darkgray176}{RGB}{176,176,176}
\definecolor{darkorange25512714}{RGB}{255,127,14}
\definecolor{lightgray204}{RGB}{204,204,204}
\definecolor{orange}{RGB}{255,165,0}
\definecolor{steelblue31119180}{RGB}{31,119,180}

\begin{axis}[
legend cell align={left},
legend style={
  fill opacity=0.8,
  draw opacity=1,
  text opacity=1,
  at={(0.03,0.03)},
  anchor=south west,
  draw=lightgray204
},
tick align=outside,
tick pos=left,
title={CelebA Denoising Results on Number of Filters},
x grid style={darkgray176},
xlabel={Number of Filters},
xmin=0.65, xmax=8.35,
xtick style={color=black},
y grid style={darkgray176},
ylabel={PSNR},
ymin=25, ymax=32,
ytick style={color=black}
]
\addplot [semithick, steelblue31119180]
table {%
1 29.12
2 29.83
3 29.79
4 29.24
5 28.78
6 28.45
7 28.21
8 27.93
};
\addlegendentry{Global Noise}
\addplot [semithick, darkorange25512714]
table {%
1 31.35
2 31.77
3 31.79
4 31.9
5 31.68
6 31.31
7 31.08
8 30.63
};
\addlegendentry{Heterogeneous Noise}
\addplot [semithick, blue, dash pattern=on 1pt off 3pt on 3pt off 3pt]
table {%
1 29.12
2 29.59
3 29.45
4 28.97
5 28.64
6 28.22
7 28.03
8 27.76
};
\addlegendentry{Global Noise (Equal Flops)}
\addplot [semithick, orange, dash pattern=on 1pt off 3pt on 3pt off 3pt]
table {%
1 31.35
2 31.66
3 31.69
4 31.74
5 31.62
6 31.07
7 30.68
8 30.23
};
\addlegendentry{Heterogeneous Noise (Equal Flops)}
\end{axis}

\end{tikzpicture}
    \caption{Denoising results of our synthesized CelebA dataset on different number of decoders. The solid lines show the results of models whose decoders are identical. On another set of experiments, we manually control the number of flops of all models to 13G, regardless of the number of decoders. We label the results as \textit{Equal Flops} in the figure.}
    \label{fig:celeba_num_filter}
\end{figure}

Second, we try to remove or modify some of the current settings and observe the results. We build the following models for comparison:
\begin{itemize}
    \item Convolutional Autoencoder (Conv AE): a convolutional autoencoder whose encoder and decoder are the same as ours. There is only one decoder and we are not applying contrastive learning to pre-train the autoencoder.
    
    \item Single Decoder PSL-AE: identical with a convolutional autoencoder with contrastive learning pretraining.
    
    \item PSL-AE (Data Aug): we originally paired up the clean and noisy patches to compute contrastive loss in the pretraining step. In this setting, instead of pairing up the clean and noisy patches, we apply data augmentations (rotation, zooming, etc.) to the noisy patches, then compute contrastive loss based on the noisy patches and their augmentations.
    
    \item PSL-AE (no KL): we train the PSL-AE using MSE and KL divergence as the reconstruction loss, as stated in Equation \ref{eq:loss:enc}. In this setting, we remove the KL loss and only use MSE in the training stage.
\end{itemize}
All the models are trained under the same settings stated in the Supplementary. Table \ref{tab:celeba_ablation} compares the denoising qualities of the models for the ablation study. We can observe that our PSL-AE still achieves the best performance on both the global noise dataset and the heterogeneous noise dataset. Therefore, we can conclude that the current settings are optimal.

\begin{table}[t]
\caption{Ablation Study on CelebA Synthesized Dataset}
\begin{center}
 \begin{tabular}{||c c c||} 
 \hline
  & PSNR (Global) & PSNR (Heter)  \\ [0.5ex] 
 \hline
 Image with Artifacts & 17.78 & 19.35 \\
 Conv AE & 28.53 & 31.02 \\
 Single Decoder PSL-AE & 29.12 & 30.15 \\
 PSL-AE (Data Aug) & 28.69 & 30.92 \\
 PSL-AE (no KL) & 29.14 & 31.25 \\
 PSL-AE & \textbf{29.59} & \textbf{31.74} \\ [1ex]
 \hline
\end{tabular}
\end{center}
\label{tab:celeba_ablation}
\end{table}

\subsection{SIDD Denoising Result}
\label{sec:sidd_experiment}
The Smartphone Image Denoising Dataset (SIDD) is a comprehensive, high-resolution image dataset derived from five emblematic smartphone cameras, capturing 10 distinct scenes. Each acquired noisy image encompasses artifacts from varying ISO levels, illumination, lighting conditions, and signal-dependent noise. Owing to the realistic scenarios under which the noise is generated, this dataset serves as an essential benchmark for evaluating the performance of denoising algorithms.

\begin{figure}[t]
\begin{center}
    \includegraphics[width=0.9\linewidth]{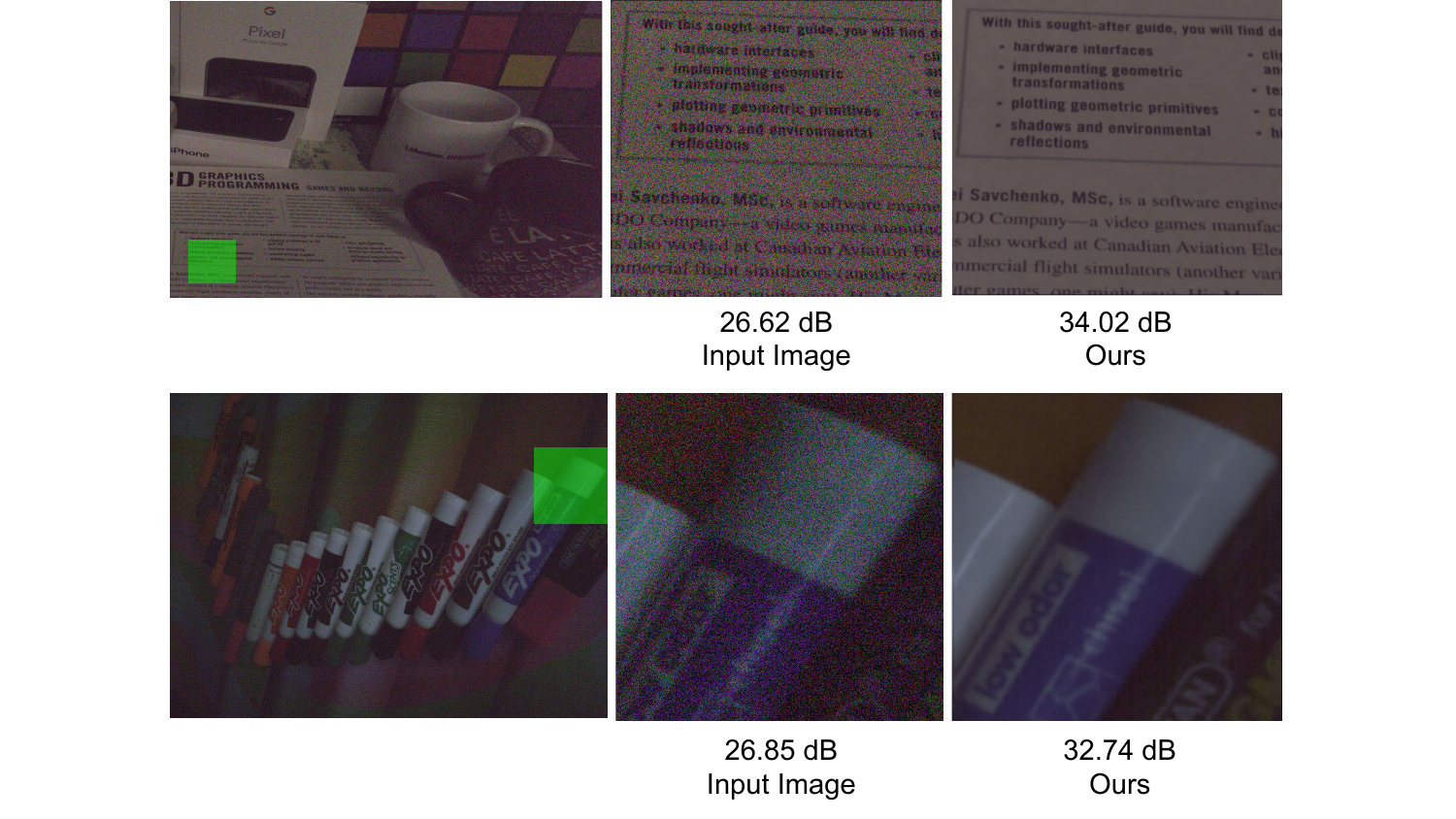}
    \includegraphics[width=0.9\linewidth]{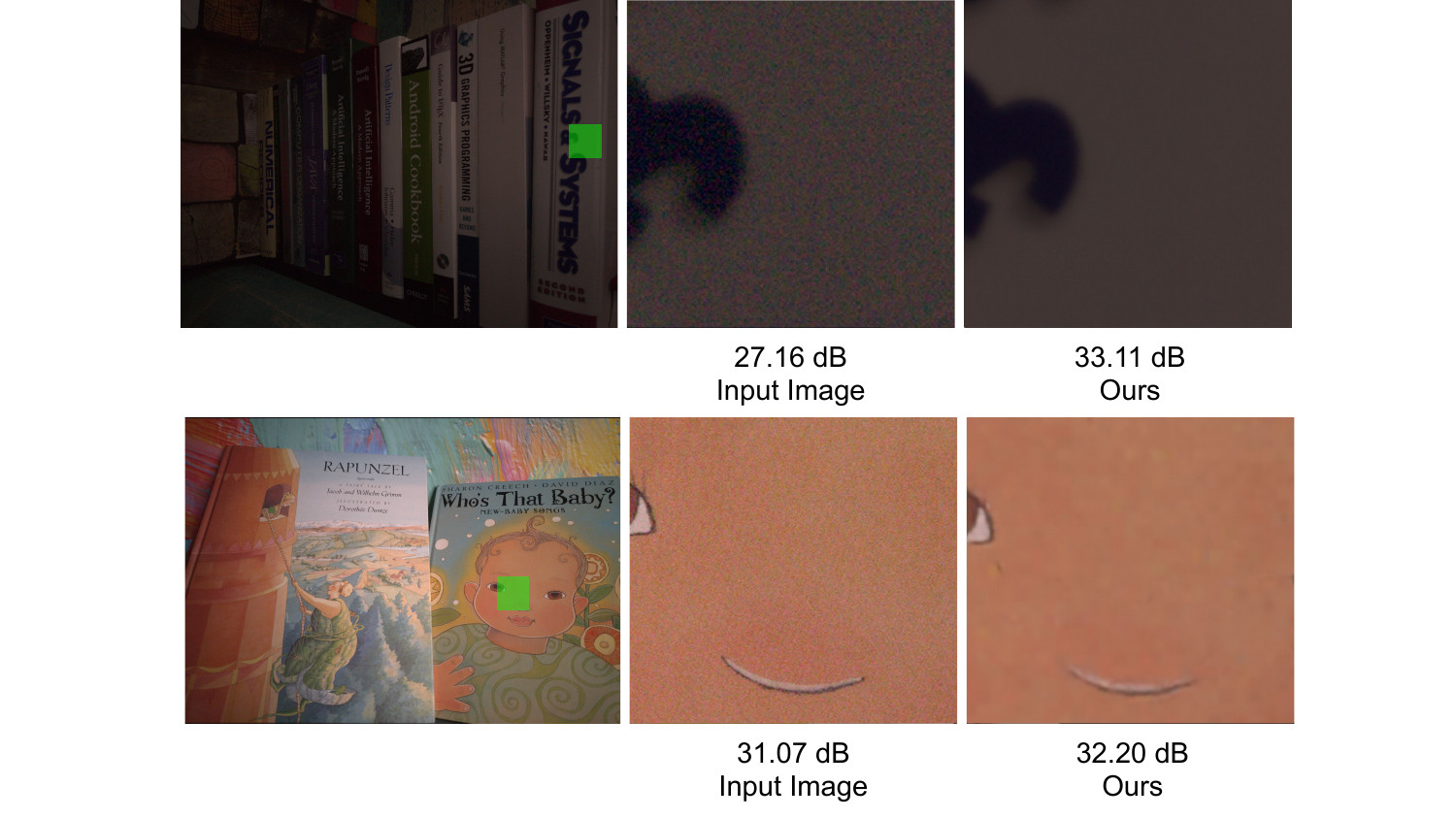}
\end{center}
\caption{SIDD denoising results. For an enhanced visualization of the denoising results, the region covered by the green rectangle within each figure in the left column is magnified. The figures sequentially presented from the left column to the right column are: comprising the original noisy images sourced from the SIDD dataset, the magnified noisy sections, and the corresponding magnified denoised segments.}
\label{fig:sidd}
\end{figure}
\paragraph{Experimental Setup}
We select 320 sRGB images from the SIDD dataset to train our PSL-AE network and use the SIDD Benchmark Data (40 images) for validation.

\paragraph{Our denoising results on SIDD}

We compare our PSL-AE with several benchmark denoising methods (BM3D, NLM, and KSVD) and novel deep learning networks (DnCNN, CBDNet, etc.) that can be used for denoising, presented in Table \ref{tab:sidd}. \emph{Noisy images} in Table \ref{tab:sidd} refer to the images before denoising procedures.
DnCNN and CBDNet are two of the deep learning networks that have a similar complexity to ours; NBNet and UNet are two of the state-of-the-art deep neural networks that can be used for image denoising. In addition to the quality metrics, we compute the number of flops of each deep learning network. The number of flops indicates the complexity of the network, as well as the potential training time.

Our model significantly outperforms those classical methods. Comparing with DnCNN and CBDNet, our network slightly outperforms them. However, our PSL-AE is not able to perform as good as the current state of the arts methods. But the complexities of the state of the arts methods are far beyond our network, hence the required training time of their networks are also significantly exceeding ours.

In addition, we observe that patches from images in SIDD tend to be assigned to a single cluster based on the same set of features they have. Therefore, there is no significant improvement using multiple decoders compared to the result in Sythesized Heterogeneous Noise CelebA dataset. Even though the proximity of noisy patterns do affect the result, we still observe a better performance of multi-decoder PSL-AE. This proves our idea that learning latent subspace is promising to the large-scale, realistic dataset, if we complicate and design the appropriate encoder/decoders.

Figure \ref{fig:sidd} presents some denoising results of our PSL-AE. We provide similar ablation study as in CelebA case and Figure \ref{fig:sidd_num_filter} shows how the number of decoders affects the outcomes of our model.

\begin{table}[t]
\caption{SIDD sRGB to sRGB Results (Small Scale)}
\begin{center}
 \begin{tabular}{||c c c c||} 
 \hline
  & PSNR & SSIM & Flops\\ [0.5ex] 
 \hline
 Noisy Image & 31.18 & 0.831 & /\\
 
 BM3D\cite{dabov2007image}& 25.65 & 0.685 & /\\
 
 NLM\cite{buades2005non}& 26.75 & 0.699 & / \\
 
 KSVD\cite{1710377} & 26.88 & 0.842 & /\\
 
 DnCNN\cite{dncnn} & 30.71 & 0.695 & 25.396G \\
 
 PSE-RL\cite{ch2021reinforcement} & 32.53 & 0.887 & 13.573G \\
 
 CBDNet\cite{guo2018convolutional} & 33.28 & 0.868 & 21.537G \\
 
 NBNet\cite{nbnet} & 39.75 & / & 152.624G \\
 
 UNet\cite{unet} & 39.62 & / & 116.385G \\

 Single Decoder Conv & 32.62 & 0.892 & 7.462G \\
 
 PSL-AE & \textbf{33.81} & \textbf{0.903} & 12.581G \\[1ex]
 \hline
\end{tabular}
\end{center}
\label{tab:sidd}
\end{table}
\begin{figure}[ht]
    \centering
\begin{tikzpicture}

\definecolor{darkgray176}{RGB}{176,176,176}
\definecolor{darkorange25512714}{RGB}{255,127,14}
\definecolor{lightgray204}{RGB}{204,204,204}
\definecolor{steelblue31119180}{RGB}{31,119,180}

\begin{axis}[
legend cell align={left},
legend style={fill opacity=0.8, draw opacity=1, text opacity=1, draw=lightgray204},
tick align=outside,
tick pos=left,
title={SIDD Denoising Results on Number of Filters},
x grid style={darkgray176},
xlabel={Number of Filters},
xmin=0.65, xmax=8.35,
xtick style={color=black},
y grid style={darkgray176},
ylabel={PSNR},
ymin=31.0485, ymax=33.9415,
ytick style={color=black}
]
\addplot [semithick, steelblue31119180]
table {%
1 33.11
2 33.76
3 33.81
4 33.59
5 33.03
6 32.68
7 32.1
8 31.83
};
\addlegendentry{Identical Decoder}
\addplot [semithick, darkorange25512714]
table {%
1 33.11
2 33.54
3 33.55
4 33.47
5 32.87
6 32.45
7 31.83
8 31.18
};
\addlegendentry{Equal Flops}
\end{axis}

\end{tikzpicture}
    \caption{Denoising results on SIDD dataset on different number of decoders. The blue line shows the results of models whose decoders are identical. The orange line shows the results of models whose total number of flops are approximately equal. The approach of controlling the total number of flops is stated in Section \ref{sec:ablation}.}
    \label{fig:sidd_num_filter}
\end{figure}
\section{Conclusion}

Our PSL-AE solution provides an  enhanced Camera ISP, in particular  with unsupervised filtering of heterogeneous image artifacts. Our single-encoder, multiple-decoder networks enable the grouping of patches from images based on the severity and type of patch image distortion. The training contains two steps: learn to provide latent subspaces for patches and learn different decoders to perform patch filtering tasks heterogeneously. We conduct comprehensive experiments over both a synthesized dataset and a state-of-the-art realistic image denoising dataset SIDD, and demonstrate that the multi-decoder structure outperforms the single-decoder structure. Though our work is limited to high-level detail preserving artifact filtering, our method should enable to generate a higher-quality image for any CameraISP. Our future work includes the extension of our network architecture to perform high-resolution image denoising, deblurring and robustness to image corruption.  

\section{Acknowledgement}
Research for this paper was supported in part by the Michael J Fox Foundation, and in part from a grant from the Army Research Office accomplished under Cooperative Agreement No. W911NF-19-2-0333. 

\bibliographystyle{IEEEtran}
\bibliography{ref}

\end{document}